\documentclass[review, sort&compress]{elsarticle}
\usepackage{epsfig,amsmath,graphicx,amssymb,overpic}
\usepackage{bm}
\usepackage{mathrsfs}
\usepackage{graphicx}
\usepackage{epsfig}
\usepackage{dcolumn}% Align table columns on decimal point
\usepackage{bm}% bold math
\usepackage{amsmath,amssymb,amsthm}
\usepackage[colorlinks=true,linkcolor=blue]{hyperref}
\usepackage{subfigure}
\usepackage{booktabs}
\usepackage[mathscr]{euscript}
\usepackage{ulem}
\usepackage{color}
\usepackage{geometry}  
\newcommand{\bea}{\begin{eqnarray}}
\newcommand{\eea}{\end{eqnarray}}
\newcommand{\beq}{\begin{equation}}
\newcommand{\eeq}{\end{equation}}
\newcommand{\arctanh}{arctanh}

\def\/{\over}

\usepackage{CJK}
\begin{document}
\begin{CJK*}{GBK}{song}	
\title{Exploring the depths of symmetry in the mKdV equation: physical interpretations and multi-wave solutions}

\author[rvt1]{Xiazhi Hao}%
\author[rvt2]{S. Y. Lou\corref{cor1}}
\cortext[cor1]{Corresponding author. Email: lousenyue@nbu.edu.cn(S. Y. Lou) }

\address[rvt1]{School of Mathematical Sciences, Zhejiang University of Technology, Hangzhou, 310014, China}
\address[rvt2]{School of Physical Science and Technology, Ningbo University, Ningbo, 315211, China}

\begin{abstract}	
	This manuscript embarks on an in-depth exploration of the modified Korteweg-de Vries (mKdV) equation, with a particular emphasis on unraveling the intricate structure of its infinite symmetries and their physical interpretations. Central to this investigation are the $K$-symmetries and $\tau$-symmetries, which are delineated by a recursive relationship and constitute  an infinite ensemble that underpins the conservation laws. 
	We engage with an existing symmetry conjecture, which posits that the currently identified symmetries represent a subset of a more expansive, yet to be unearthed, set. 
	This conjecture is substantiated through an analysis of the soliton solutions associated with the mKdV equation, demonstrating that these symmetries can be decomposed into linear combinations of center and wave number translation symmetries. 
	Further, by imposing an infinite sequence of symmetry constraints, it becomes feasible to derive exact multi-wave solutions.
	 This methodology, predicated on the proposed symmetry conjecture, facilitates the extraction of exact solutions, encompassing complexiton, breather, multi-soliton solutions, among others. 
\end{abstract}

%\pacs{05.45.Yv,02.30.Ik,47.20.Ky,52.35.Mw,52.35.Sb}
\maketitle
\large
\section{Introduction}

The modified Korteweg-de Vries (mKdV) equation, a cornerstone in the field of integrable systems, has been pivotal in understanding the behavior of nonlinear dispersive waves. This equation extends its relevance to a myriad of physical phenomena, including shallow water waves \cite{mbib23}, nonlinear optics \cite{hbib15}, plasma physics \cite{abib35} and so on. The mKdV equation is particularly noted for its soliton solutions \cite{sbib52}, stable, localized waves that maintain their form while propagating at a constant velocity, and its rich algebraic structure, which encapsulates an infinite number of symmetries \cite{pbib7,ybib21,sbib15,abib36} and conservation laws. The exploration of these solutions and their underlying symmetries is crucial for deciphering the complex dynamics of nonlinear phenomena.

The quest to elucidate the physical essence of the symmetries and conservation laws intrinsic to integrable systems, exemplified by the mKdV equation, represents a formidable intellectual endeavor situated at the confluence of physics and mathematics. 
The seminal contributions of Miura, Gardner, and Kruskal \cite{rbib23}, complemented by the subsequent scholarly work of \cite{pbib7,tbib3} and Fuchssteiner \cite{wbib14,abib17,bbib5,bbib6,bbib7}, have unveiled the potential for an infinite number of symmetries within integrable systems.
While these symmetries have been instrumental in tackling nonlinear physical phenomena, a cadre of foundational inquiries remain unresolved, necessitating a renewed scrutiny of their physical significance and mathematical integrity.

This manuscript embarks on an in-depth investigation into the physical interpretation related to the boundless symmetries within integrable systems, with a particular focus on the mKdV equation. We delve into the multi-wave solutions of the mKdV equation, employing an analytical methodology to scrutinize their symmetries. Our investigation is prompted by the recognition that while integrable systems are characterized by an infinite number of symmetries and conservation laws, only a select few possess explicit, tangible physical interpretations, such as those related to space-time translational invariance, Galilean or Lorentz invariance, scaling invariance, as well as the conservation of mass, momentum, and energy. We aim to bridge this gap by examining the symmetries in the context of multi-wave solutions, with a specific emphasis on the $n$-soliton solution.

The structure of this manuscript proceeds as delineated: after an introduction to the mKdV equation and its significance, we begin by providing an intrinsic physical interpretation of the infinite $K$-symmetries and $\tau$-symmetries \cite{wbib28} inherent in the mKdV equation. Subsequent section will address the implications of these symmetries for the derivation of multi-wave solutions. We introduce a novel approach to solving multi-wave solutions of the mKdV equation by harnessing the power of generalized local symmetries. Finally, we conclude with a synthesis of our findings and discuss the implications of our results.

\section{Intrinsic physical interpretations of the infinite $K$-symmetries and $\tau$-symmetries inherent in the mKdV equation}

The mKdV equation, a quintessential partial differential equation, is articulated as follows
\begin{eqnarray}\label{mkdv}
u_t=u_{xxx}+6u^2u_x.
\end{eqnarray} 
This equation is pivotal in the study of integrable systems and is distinguished by its capacity to support soliton solutions, as well as its rich algebraic structure, which encapsulates the dynamics of wave phenomena.
The infinitesimal symmetries of the mKdV equation are characterized by their invariance under the linearized mKdV equation
\begin{eqnarray}\label{syme}
\sigma_t=\sigma_{xxx}+12uu_x\sigma+6u^2\sigma_x.
\end{eqnarray}
This definition embodies the notion that the mKdV equation maintains its form under the infinitesimal transformation $u \rightarrow u+\epsilon \sigma$ with $\epsilon$ being an infinitesimal parameter.
The mKdV equation is distinguished by an infinite sequence of symmetries, known as the $K$-symmetries and $\tau$-symmetries, which are defined in a recursive manner. 
The $K$-symmetries, which are independent of the spatial and temporal variables, are articulated through the operator $\Phi$, given by
\begin{eqnarray}
\Phi\equiv \partial_x^2+4u^2+4u_x\partial_x^{-1}u.
\end{eqnarray}
The symmetries themselves are explicitly defined as \cite{pbib7}
\begin{eqnarray}
&&K_{n+1}=\Phi^n u_x, 
~~ n=0,1,...,\infty.
\end{eqnarray}
The initial members of this recursive sequence are explicitly enumerated as 
\begin{eqnarray}\label{ks}
\begin{split}
K_1&=u_x,\\
K_2&=(u_{xx}+2u^3)_x=u_t,\\
K_3&=(u_{4x}+10u^2u_{xx}+10uu_x^2+6u^5)_x,\\
K_4&=[u_{6x}+14u^2u_{4x}+56uu_xu_{xxx}+70(u^4+u_x^2)u_{xx}+42uu_{xx}^2+140u^3u_x^2+20u^7]_x,\\
K_5&=[u_{8x}+18u^2u_{6x}+108uu_xu_{5x}+6(21u^4+38uu_{xx}+35u_x^2)u_{4x}+138uu_{xxx}^2\\&+4(252u^3+189u_{xx})u_xu_{xxx}+182u_{xx}^3+756u^3u_{xx}^2+84(5u^4+37u_x^2)u^2u_{xx}\\&+798uu_x^4+1260u^5u_x^2+70u^9]_x.
\end{split}
\end{eqnarray}
The $\tau$-symmetries, which  depend explicitly on the
spatial and temporal variables, are given by \cite{ybib21}
\begin{eqnarray}
&&\tau_{n+1}=\Phi^n(xu_x+3tu_t+u),~~ n=0,1,...,\infty.
\end{eqnarray}
The initial entries of this sequence are explicitly presented as
\begin{eqnarray}\label{ts}
\begin{split}
\tau_1&=xu_x+3tu_t+u=xK_1+3tK_2+u,\\
\tau_2&=xK_2+3tK_3+2K_1\int u^2 {\rm d}x+3K_{1,x}+4u^3,\\
\tau_3&=xK_3+3tK_4+2K_2\int u^2 {\rm d}x+6K_1\int u^4 {\rm d}x-6K_1\int K_1^2 {\rm d}x+10u^2K_{1,x}+5K_{2,x}+16u^5,\\
\tau_4&=xK_4+3tK_5+2K_3\int u^2 {\rm d}x+6K_2\int u^4 {\rm d}x+20K_1\int u^6 {\rm d}x-6K_2\int K_1^2 {\rm d}x\\&+10K_1\int K_{1,x}^2 {\rm d}x-100K_1\int u^2K_1^2 {\rm d}x-6K_1^2K_{1,x}+14u^2K_{2,x}+42u^4K_{1,x}+7K_{3,x}+64u^7.
\end{split}
\end{eqnarray} 
 The symmetries inherent to the mKdV equation act as the progenitors of an infinite sequence of conservation laws, which are fundamental in the study of integrable systems. These conservation laws are articulated through a series of differential identities, with the initial members of this sequence delineated as \cite{rbib24}
\begin{eqnarray}\label{con1}
&&u_t=(2u^3+u_{xx})_x,\\\label{con2}
&&\left(u^2\right)_t=(3u^4+2uu_{xx}-u_x^2)_x,\\\label{con3}
&&\left(u^4-u_x^2\right)_t=\left(4u^6+4u^3u_{xx}-12u^2u_x^2-2u_xu_{xxx}+u_{xx}^2\right)_x,\\\label{con4}
&&\left(u^6-5u^2u_x^2+\frac{u_{xx}^2}{2}\right)_t=\left(\frac92 u^8+6u^5u_{xx}-45u^4u_x^2+8u^2u_{xx}^2-10u^2u_xu_{xxx}\right.\\\nonumber&&\left.+10uu_x^2u_{xx}+\frac{u_x^4}{2}+u_{xx}u_{4x}-\frac{u_{xxx}^2}{2}\right)_x.
\end{eqnarray}
Herein, the conservation of mass is manifested as the mKdV equation itself \eqref{con1}, expressed in a conservation form. The conservation of momentum emerges as the second conserved quantity \eqref{con2}, derived by multiplying the mKdV equation by $2u$ and subsequently reformulating it in conservation form. The conservation of energy, represented by the third conserved quantity \eqref{con3}, is more intricate to derive, necessitating a series of multiplications and factorizations.
Beyond these initial three conserved densities, subsequent members of the sequence do not readily lend themselves to a straightforward physical interpretation, suggesting a potential avenue for further theoretical inquiry and elucidation. %The exploration of these higher conserved densities and their physical underpinnings remains an open and challenging problem in the analysis of integrable systems such as the mKdV equation.

Transitioning into the realm of differential geometry and the theory of partial differential equations, 
a Lie point symmetry is rigorously characterized by a vector field of the form
\begin{eqnarray}\label{sym}
\sigma=U(x,t,u)-X(x,t,u)u_x-T(x,t,u)u_t,
\end{eqnarray}
where $U, X$ and $T$ are functions that characterize the symmetry transformation.
Upon immersion of this ansatz into the linearized mKdV symmetry equation and subsequent elimination of $u_t$
by the mKdV equation itself, the resultant Lie point symmetry for the mKdV equation is derived to be
\begin{eqnarray*}
	\sigma&=&cu+(cx+x_0)u_x+(3ct+t_0)u_t\\
	&=&c(u+xu_x+3tu_t)+x_0u_x+t_0u_t\\
	&=&c\tau_1+x_0K_1+t_0K_2,
\end{eqnarray*}
where $c, x_0$ and $t_0$ are arbitrary constants, $K_1$ and $K_2$ correspond to the spatial and temporal translational symmetries, respectively. $\tau_1$ encapsulates the scaling symmetry, reflecting the homothetic transformation of the solution space.
In contradistinction to the KdV equation, which is endowed with Galilean invariance, the mKdV equation does not partake in this property. Nonetheless, the mKdV equation does exhibit form invariance under the specified translational and scaling transformations. 
Accordingly, the mKdV equation \eqref{mkdv} exhibits invariance for a specified solution $ u = u_0(x,t)$ under the operations of spatial and temporal translations, as well as under appropriate scaling transformations. Specifically, for any real numbers $ x_0 $ and $ t_0 $, and for any positive constant $ c$, the transformations $ u_0(x - x_0, t - t_0) $ and $ e^{c\epsilon}u_0(e^{c\epsilon}x, e^{3c\epsilon}t) $ are solutions of the mKdV equation \eqref{mkdv}, thereby demonstrating its form invariance.
The challenge of augmenting the solution $u= u_0(x, t)$ with additional arbitrary constants through the utilization of higher order $K$-symmetries and $\tau$-symmetries remains an unresolved issue.

To elucidate the potential physical interpretations, it is imperative to specify the exact form of the solution $u_0(x, t)$ for the mKdV equation. The mKdV equation is renowned for its diverse soliton solutions, among which the $n$-soliton solution stands out as a fundamental and significant case, embodying a particular structure as documented in \cite{sbib52}
\begin{eqnarray}\label{ns}
u_{ns}=\pm {\rm i} \left(\ln \frac{F_{-}}{F_{+}}\right)_x,
\end{eqnarray}
where the functions $F_{\pm}$ are defined by the sum over all permutations $\mu_j=0,1$ for $j=1,2,...,n$, and are given by
\begin{eqnarray}
F_{\pm}=\sum_{\mu=1,0}\exp\left[\sum_{j=1}^{n}\mu_j\left(\xi_j\pm \frac{{\rm i}\pi}{2}\right)+\sum_{1\leq j<l}^n\mu_j\mu_l\theta_{jl}\right].
\end{eqnarray}
The terms within the exponent are specified by
\begin{eqnarray}
\xi_j=k_jx+k_j^3t+c_{j}, ~~\exp(\theta_{jl})=\left(\frac{k_j-k_l}{k_j+k_l}\right)^2.
\end{eqnarray}
\\
In this formulation, $\xi_j$ represents the phase shift of the $j$-th soliton, with $k_j$ being its wave number and $c_{j}$ a constant of integration. The quantity $\theta_{jl}$ captures the interaction between the $j$-th and $l$-th solitons, encapsulating the essence of their mutual dynamics.

From the expression delineated by equation \eqref{ns}, it is observed that the solution exhibits invariance under the transformation 
\begin{eqnarray}\label{tf}
F_{\pm}\rightarrow \beta \exp(Kx+\Omega t+X_0)F_{\pm}
\end{eqnarray}
with arbitrary constants $K, \beta, \Omega$ and $X_0$.
Leveraging the invariance property expressed by equation \eqref{tf}, coupled with the freedom to assign values to $c_{j}$ permits a re-expression of the solution \eqref{ns} in an alternative form
\begin{eqnarray}\label{uns}
u_{ns}=\pm 2\left(\arctan \frac{\sum_{\nu_e}K_{\nu}\sinh(\sum_{j=1}^N\nu_j\eta_j)}{\sum_{\nu_o}K_{\nu}\cosh(\sum_{j=1}^N\nu_j\eta_j)}\right)_x.
\end{eqnarray}
In this formulation, the summation over \( \nu_o \) is to be conducted for the ensemble of non-dual odd permutations of \( \nu_i \) taking values of \( \pm 1 \) for each index \( i = 1, 2, \ldots, N \), with the specific stipulation that the count of \( \nu_i = 1 \) must be odd. Concurrently, the summation over \( \nu_e \) is to be executed for the collection of non-dual even permutations of \( \nu_i \) within the same set of values, ensuring that the quantity of \( \nu_i = 1 \) is even. 
The coefficient $K_{\nu}$
is defined by the product
\begin{eqnarray}
K_{\nu}\equiv \prod_{i>j}(k_i-\nu_i\nu_jk_j),
\end{eqnarray}
which encapsulates the pairwise interactions between different components of the solution. The term $\eta_j$
is defined as
\begin{eqnarray}
\eta_j=\frac12 \xi_j=\frac12 (k_jx+k_j^3t+c_j),
\end{eqnarray}
representing a phase shift that incorporates spatial and temporal dependencies alongside a constant phase $c_j$.
\iffalse
The solutions for one-, two-, and three-soliton scenarios are presented in a simplified form
\begin{eqnarray}
u_{1s}=\pm 2\left[\arctan\left(\frac{\sinh(\eta_1)}{\cosh(\eta_1)}\right)\right]_x,
\end{eqnarray}
\begin{eqnarray}
u_{2s}=\pm 2\left[\arctan\left(\frac{(k_1-k_2)\sinh(\eta_1+\eta_2)}{(k_1+k_2)\cosh(\eta_1-\eta_2)}\right)\right]_x.
\end{eqnarray}
%3孤子不对???
The three-soliton solution captures the complex interaction among three solitons, with the solution expressed as
\begin{eqnarray}
u_{3s}=\pm 2\left[\arctan \frac{A}{B}\right]_x,
\end{eqnarray}
where $A$ and $B$ are functions of the wave numbers $k_1,k_2$ and $k_3$, as well as the phases $\eta_1,\eta_2$ and $\eta_3$
\begin{eqnarray*}
	A&=&(k_2-k_1)(k_3-k_1)(k_3-k_2)\sinh(-\eta_1-\eta_2-\eta_3)\\&+&(k_2-k_1)(k_3+k_1)(k_3+k_2)\sinh(\eta_1+\eta_2-\eta_3)\\&+&(k_2+k_1)(k_3-k_1)(k_3+k_2)\sinh(\eta_1-\eta_2+\eta_3)\\&+&(k_2+k_1)(k_3+k_1)(k_3-k_2)\sinh(\eta_1+\eta_2-\eta_3),\\
	B&=&(k_2-k_1)(k_3-k_1)(k_3-k_2)\cosh(\eta_1+\eta_2+\eta_3)\\&+&(k_2-k_1)(k_3+k_1)(k_3+k_2)\cosh(-\eta_1-\eta_2+\eta_3)\\&+&(k_2+k_1)(k_3-k_1)(k_3+k_2)\cosh(-\eta_1+\eta_2-\eta_3)\\&+&(k_2+k_1)(k_3+k_1)(k_3-k_2)\cosh(-\eta_1-\eta_2+\eta_3).
\end{eqnarray*}
 The expressions for $A$ and $B$ are constructed to account for all pairwise interactions among the three solitons
\fi
The solutions pertaining to the one-, two-, and three-soliton scenarios are articulated in a concise form, reflecting the fundamental structures of soliton interactions within the mKdV equation.
%1. One-Soliton solution:
The univariate soliton solution, embodying a solitary wave disturbance, is represented by the formula
\begin{eqnarray}\label{sol1}
u_{1s} = \pm 2 \left[ \arctan \left( \frac{\sinh(\eta_1)}{\cosh(\eta_1)} \right) \right]_x,
\end{eqnarray}
where the argument of the arctan function captures the ratio of the hyperbolic sine to cosine of the phase $\eta_1$.
%2. Two-Soliton solution:
The bivariate soliton solution, detailing the interaction between two solitons, is given by
\begin{eqnarray}\label{sol2}
u_{2s} = \pm 2 \left[ \arctan \left( \frac{(k_1 - k_2) \sinh(\eta_1 + \eta_2)}{(k_1 + k_2) \cosh(\eta_1 - \eta_2)} \right) \right]_x,
\end{eqnarray}
with $k_1$ and $k_2$ denoting the wave numbers of the two solitons and their interaction encoded in the hyperbolic functions of their combined phases $\eta_1$ and $\eta_2$.
%3. Three-Soliton solution:
The trivariate soliton solution, encapsulating the intricate dynamics among three solitons, is expressed as
\begin{eqnarray}\label{sol3}
u_{3s} = \pm 2 \left[ \arctan \frac{A}{B} \right]_x,
\end{eqnarray}
where $A$ and $B$ are functions that embody the collective interactions influenced by the wave numbers $k_1$, $k_2$, and $k_3$, and the phases $\eta_1$, $\eta_2$, and $\eta_3$. The expressions for $A$ and $B$ are meticulously constructed to incorporate all pairwise interactions among the solitons, and are defined as follows:
\begin{eqnarray*}
	A&=&(k_2-k_1)(k_3-k_1)(k_3-k_2)\sinh(-\eta_1-\eta_2-\eta_3)\\&+&(k_2-k_1)(k_3+k_1)(k_3+k_2)\sinh(\eta_1+\eta_2-\eta_3)\\&+&(k_2+k_1)(k_3-k_1)(k_3+k_2)\sinh(\eta_1-\eta_2+\eta_3)\\&+&(k_2+k_1)(k_3+k_1)(k_3-k_2)\sinh(\eta_1+\eta_2-\eta_3),\\
	B&=&(k_2-k_1)(k_3-k_1)(k_3-k_2)\cosh(\eta_1+\eta_2+\eta_3)\\&+&(k_2-k_1)(k_3+k_1)(k_3+k_2)\cosh(-\eta_1-\eta_2+\eta_3)\\&+&(k_2+k_1)(k_3-k_1)(k_3+k_2)\cosh(-\eta_1+\eta_2-\eta_3)\\&+&(k_2+k_1)(k_3+k_1)(k_3-k_2)\cosh(-\eta_1-\eta_2+\eta_3).
\end{eqnarray*}
The expressions for $A$ and $B$ are constructed to account for all pairwise interactions among the three solitons.

For the explicitly determined solution $u=u_0=u_{ns}$, as characterized by equation \eqref{ns}, there exists a complement of $2n$ symmetries that are physically pertinent. These symmetries are rigorously defined within the mathematical structure of the equation as follows
\begin{eqnarray}
\sigma_{c_m}=u_{ns,c_m}=\pm {\rm i}\left(\frac{F_{-,c_m}}{F_-}-\frac{F_{+,c_m}}{F_+}\right)_x,
\end{eqnarray} 
where the functions $ F_{\pm,c_m}$ are defined by the summation over the binary variable $\mu$ and given by
\begin{eqnarray*}
F_{\pm,c_m}=\sum_{\mu=1,0}\mu_m \exp\left[\sum_{j=1}^{n}\mu_j\left(\xi_j\pm \frac{{\rm i}\pi}{2}\right)+\sum_{1\leq j<l}^n\mu_j\mu_l\theta_{jl}\right], m=1,2,...,n.
\end{eqnarray*}
Similarly, the symmetries denoted by $\sigma_{k_m}$
are formulated as
\begin{eqnarray}
\sigma_{k_m}=u_{ns,k_m}=\pm {\rm i}\left(\frac{F_{-,k_m}}{F_-}-\frac{F_{+,k_m}}{F_+}\right)_x,
\end{eqnarray}
with the associated functions $F_{\pm,k_m}$ being
\begin{eqnarray*}
F_{\pm,k_m}=\sum_{\mu=1,0}\left(\mu_m\xi_{m,k_m}+\sum_{1\leq i<p}^n\mu_i\mu_p\theta_{ip,k_m} \right)\exp\left[\sum_{j=1}^{n}\mu_j\left(\xi_j\pm \frac{{\rm i}\pi}{2}\right)+\sum_{1\leq j<l}^n\mu_j\mu_l\theta_{jl}\right].
\end{eqnarray*}
The symmetries $ \sigma_{c_m} $ and $ \sigma_{k_m} $ encapsulate the translational invariance properties of the $ m$-th soliton within the solution framework. Specifically, 
the symmetry  $ \sigma_{c_m} $ corresponds to the central translational invariance of the $ m$-th soliton, reflecting the capacity of the soliton to maintain its structural integrity upon spatial translation. Conversely, $ \sigma_{k_m} $ encapsulates the wave number translational invariance of the $m$-th soliton, where the wave number $ k_m $ is intricately linked to the amplitude, spatial extent, and propagation velocity, thus embodying the dynamic characteristics of the $m$-th soliton.

In light of these conceptualizations, a profound and pertinent question arises: what are the intrinsic connections, if any, between the higher-order symmetries $ \{ K_{n+1}, \tau_{n+1} \} $ and the sets of translational invariants $ \{ \sigma_{c_m}, \sigma_{k_m} \} $?

For the $K$-symmetries as delineated by system \eqref{ks}, with the solution $u = u_{ns}$ as specified by equation \eqref{ns}, the derivation of the corresponding expressions is a matter of straightforward computation
\begin{eqnarray}
\left. K_1\right|_{u=u_{ns}}&=&\left.u_x\right|_{u=u_{ns}}
=\sum_{m=1}^n k_m u_{ns,c_m}=\sum_{m=1}^n k_m\sigma_{c_m}, \nonumber\\
\left. K_2\right|_{u=u_{ns}}&=&\left.\Phi u_x\right|_{u=u_{ns}}
=\sum_{m=1}^n k_m^{3} u_{ns,c_m}=\sum_{m=1}^n k_m^{3}\sigma_{c_m},\nonumber\\
\left. K_3\right|_{u=u_{ns}}&=&\left.\Phi^{2}u_x\right|_{u=u_{ns}}
=\sum_{m=1}^n k_m^{5} u_{ns,c_m}=\sum_{m=1}^n k_m^{5}\sigma_{c_m},\nonumber\\
&  \vdots &\nonumber\\
\left. K_i\right|_{u=u_{ns}}&=&\left.\Phi^{i-1}u_x\right|_{u=u_{ns}}
=\sum_{m=1}^n k_m^{2i-1}u_{ns,c_m}=\sum_{m=1}^n k_m^{2i-1}\sigma_{c_m},\
i=4,\ 5,\ \ldots, \ \infty. \label{Kicm}
\end{eqnarray}This sequence explicitly demonstrates that,  within the purview of the specified solution \eqref{ns}, the infinite series of $K $-symmetries lacks inherent autonomy. Precisely, only $n $ of these symmetries, characterized by $ K_i $ for $ i = 1, \ldots, n $, are independent. The subsequent symmetries, which are unbounded in number, are revealed to be linearly reducible to these independent symmetries
\begin{eqnarray}
&&\left. K_i\right|_{u=u_{ns}}=\left.\Phi^{i-1}u_x\right|_{u=u_{ns}}
=\left.\sum_{m=1}^n k_m^{2i-1}\frac{\Delta_{nm}}{\Delta_n}\right|_{u=u_{ns}},\
i=n+1,\ n+2,\ \ldots, \ \infty. \label{Kcm}
\end{eqnarray}
Here, $\Delta_n$ is utilized to denote the determinant of the $n \times n$ matrix $ M $, which is formally defined as $ \Delta_n \equiv \det(M) $. The matrix $ M $ is constituted as
\begin{eqnarray}\label{M}
M = \begin{pmatrix}
k_1 & k_2 & \cdots & k_m & \cdots & k_{n-1} & k_n \\
k_1^3 & k_2^3 & \cdots & k_m^3 & \cdots & k_{n-1}^3 & k_n^3 \\
\vdots & \vdots & \ddots & \vdots & \ddots & \vdots & \vdots \\
k_1^{2n-1} & k_2^{2n-1} & \cdots & k_m^{2n-1} & \cdots & k_{n-1}^{2n-1} & k_n^{2n-1}
\end{pmatrix},
\end{eqnarray}
where each entry $k_i$ represents a specific wave number associated with the $ i$-th soliton, and the powers denote the successive iterations of the wave number raised to the $ (2i-1) $-th power.
Conjointly, $ \Delta_{nm}$ symbolizes the determinant of the $ n \times n $ matrix $M_m$, which is defined as $ \Delta_{nm} \equiv \det(M_m) $. The matrix $ M_m $ is structured similarly to $ M $, yet it is distinguished by the replacement of the $m $-th column with the sequence $K_1, K_2, \ldots, K_n $, which are related to the $K $-symmetries of the system
\begin{eqnarray}\label{Mm}
M_m = \begin{pmatrix}
k_1 & k_2 & \cdots & K_1 & \cdots & k_{n-1} & k_n \\
k_1^3 & k_2^3 & \cdots & K_2 & \cdots & k_{n-1}^3 & k_n^3 \\
\vdots & \vdots & \ddots & \vdots & \ddots & \vdots & \vdots \\
k_1^{2n-1} & k_2^{2n-1} & \cdots & K_n & \cdots & k_{n-1}^{2n-1} & k_n^{2n-1}
\end{pmatrix}.
\end{eqnarray}
These determinants, $\Delta_n $ and $ \Delta_{nm} $, are pivotal in the analysis of the linear dependencies among the $K$-symmetries and the construction of higher order symmetries from the independent set.

In the simplest case where $u_{ns} = u_{1s}$ \eqref{sol1}, the $K$-symmetries are generated by a single parameter $k_1$. The recurrence relation simplifies to
	\begin{eqnarray}
	K_{i\geq 2}=k_1^{2i-2}K_1.
	\end{eqnarray}
	This indicates that all higher-order $K$-symmetries are simply powers of $k_1$ multiplied by the initial $K_1$.
	%​This is a straightforward case where the symmetry is entirely determined by one parameter.
	When $u_{ns} = u_{2s}$ \eqref{sol2}, the situation becomes more complex, involving two parameters $k_1$ and $k_2$. The $K$-symmetries for $i\geq 3$ are given by
\begin{eqnarray}
K_{i\geq 3}=\left(\frac{k_1^2k_2^{2i-2}-k_2^2k_1^{2i-2}}{k_1^2-k_2^2}\right)K_1+\left(\frac{k_1^{2i-2}-k_2^{2i-2}}{k_1^2-k_2^2}\right)K_2.
\end{eqnarray}
This expression shows that the higher-order $K$-symmetries are linear combinations of $K_1$ and $K_2$, with coefficients that depend on the powers of $k_1$ and $k_2$.

In the case of a three-soliton solution, with \( u_{ns} = u_{3s} \) \eqref{sol3}, the system possesses a finite set of three independent \( K \)-symmetries. These are specifically \( \left. K_1 \right|_{u = u_{3s}} \), \( \left. K_2 \right|_{u = u_{3s}} \), and \( \left. K_3 \right|_{u = u_{3s}} \). All subsequent \( K \)-symmetries are not independent but are, instead, linear combinations of these initial three, as dictated by the relationship established in equation \eqref{Kcm}
\begin{eqnarray*}\nonumber
K_{i\geq4}&=&\left(\frac{k_2^2k_3^2k_1^{2i-2}}{(k_1^2-k_2^2)(k_1^2-k_3^2)}-\frac{k_1^2k_3^2k_2^{2i-2}}{(k_1^2-k_2^2)(k_2^2-k_3^2)}+\frac{k_1^2k_2^2k_3^{2i-2}}{(k_1^2-k_3^2)(k_2^2-k_3^2)}\right)K_1\\&+&\left(-\frac{(k_2^2+k_3^2)k_1^{2i-2}}{(k_1^2-k_2^2)(k_1^2-k_3^2)}+\frac{(k_1^2+k_3^2)k_2^{2i-2}}{(k_1^2-k_2^2)(k_2^2-k_3^2)}-\frac{(k_1^2+k_2^2)k_3^{2i-2}}{(k_1^2-k_3^2)(k_2^2-k_3^2)}\right)K_2\\&+&\left(\frac{k_1^{2i-2}}{(k_1^2-k_2^2)(k_1^2-k_3^2)}-\frac{k_2^{2i-2}}{(k_1^2-k_2^2)(k_2^2-k_3^2)}+\frac{k_3^{2i-2}}{(k_1^2-k_3^2)(k_2^2-k_3^2)}\right)K_3.
\end{eqnarray*}
\iffalse
\begin{eqnarray}\nonumber
K_{i\geq4}&=&\frac{k_1^{2i-2}(k_2^2k_3^2K_1-(k_2^2+k_3^2)K_2+K_3)}{(k_1^2-k_2^2)(k_1^2-k_3^2)}-\frac{k_2^{2i-2}(k_1^2k_3^2K_1-(k_1^2+k_3^2)K_2+K_3)}{(k_1^2-k_2^2)(k_2^2-k_3^2)}\\&+&\frac{k_3^{2i-2}(k_1^2k_2^2K_1-(k_1^2+k_2^2)K_2+K_3)}{(k_1^2-k_3^2)(k_2^2-k_3^2)}.
\end{eqnarray}\fi
For the sake of notational conciseness, the  \( K \)-symmetries,  \( \left. K_i \right|_{u = u_{ns}} \) for \( i = 1, 2, \ldots \), have been uniformly represented as \( K_i \) throughout the presentation. 

It is plausible to hypothesize that the $\tau$-symmetries, as defined by equation \eqref{ts}, for the specific solution $u=u_{ns}$ detailed in equation \eqref{ns}, can be represented as linear combinations of the wave number translational symmetries $\sigma_{k_i}$ and the center translational symmetries $\sigma_{c_i}$.
Focusing on the one-soliton solution $u_{1s}$, given by equation \eqref{sol1}, we derive the following sequence of relationships
\begin{eqnarray*}
\tau_1&=&k_1\sigma_{k_1},\\
\tau_2&=&k_1^3\sigma_{k_1}-2k_1^2\sigma_{c_1}=k_1^2\tau_1-2k_1K_1,\\
\tau_3&=&k_1^5\sigma_{k_1}-4k_1^4\sigma_{c_1}=k_1^4\tau_1-4k_1^3K_1,\\
\vdots
\end{eqnarray*}
The recursive formula for the higher-order $\tau$-symmetries in relation to the set $\{\sigma_{k_1}, \sigma_{c_1}\}$, or equivalently $\{\tau_1, K_1\}$, is
\begin{eqnarray}
\tau_{i\geq 2}&=&(2-2i)k_1^{2n-3}K_1+k_1^{2i-2}\tau_1.
\end{eqnarray}

Employing the two-soliton solution $u_{2s}$ \eqref{sol2} as another example, the following expressions are obtained
\begin{eqnarray*}
\tau_1&=&k_1\sigma_{k_1}+k_2\sigma_{k_2},\nonumber\\
\tau_2&=&k_1^3\sigma_{k_1}+k_2^3\sigma_{k_2}-2k_1(k_1+2k_2)\sigma_{c_1}-2k_2(2k_1+k_2)\sigma_{c_2},\nonumber\\
\tau_3&=&k_1^5\sigma_{k_1}+k_2^5\sigma_{k_2}-4k_1(k_1^3+k_1^2k_2+k_2^3)\sigma_{c_1}
-4k_2(k_1^3+k_1k_2^2+k_2^3)\sigma_{c_2}
\nonumber\\
&=&-k_1^2k_2^2\tau_1+(k_1^2+k_2^2)\tau_2+2(k_1+k_2)(k_1k_2 K_1-K_2),\\
\vdots
\end{eqnarray*}
The recursive relationships among the higher-order $\tau$-symmetries and the set $\{\tau_1, \tau_2, K_1, K_2\}$ are given by
\begin{eqnarray*}
\tau_{i\geq 3}&=&2\left(\frac{i(k_1^{2i-3}k_2^2-k_1^2k_2^{2i-3})}{k_1^2-k_2^2}+\frac{k_1k_2^{2i-3}(k_1^2+k_1k_2-k_2^2)+k_1^{2i-3}k_2(k_1^2-k_1k_2-k_2^2)}{(k_1-k_2)(k_1+k_2)^2}\right)K_1\\
&+&2\left(\frac{k_1^{2i-3}((1-i)k_2+(2-i)k_1)+k_2^{2i-3}((i-1)k_1+(i-2)k_2)}{(k_1-k_2)(k_1+k_2)^2}\right)K_2\\
&+&\left(\frac{k_1^2k_2^{2i-2}-k_2^2k_1^{2i-2}}{k_1^2-k_2^2}-\right)\tau_1+\left(\frac{k_1^{2i-2}-k_2^{2i-2}}{k_1^2-k_2^2}\right)\tau_2.
\end{eqnarray*}

\section{New way to solve $n$-wave solution}

The resolution of similarity solutions through Lie point symmetries is a recognized approach within the mathematical physics community. These symmetries, when applied to integrable models, facilitate the discovery of diverse solution types \cite{sbib52,pbib16}. 
In this section, we delve into the application of generalized $K$-symmetries, complemented by the symmetry conjecture articulated by Lou \cite{lou1}, to extract multi-wave solutions from the mKdV equation. 
The conjecture put forth by Lou \cite{lou1} is of particular significance as it provides a framework for understanding the symmetries inherent in nonlinear partial differential systems that are symmetry integrable. It posits that for an $n$-wave solution of such a system, the dependencies and interactions between wave numbers and other parameters, like Riemann invariants, can be harnessed to uncover deeper symmetries within the system.
The crux of the conjecture lies in the assertion that the multitude of $K$-symmetries and $\tau$-symmetries, which are generated through the action of the recursion operator $\Phi$, can be decomposed into linear combinations of parameter translation symmetries. These include translations in the center of the solution, as well as translations in the wave number domain, symbolized by the partial derivatives $u_{c_i}$ and $u_{k_i}$, respectively. This decomposition reveals a rich tapestry of symmetries that are intrinsic to the solution structure, offering insights into the underlying geometry of the solution space.
By integrating symmetry conjecture with the generalized $K$-symmetries, we are thus able to construct a comprehensive symmetry group that encapsulates the multi-wave solutions of the mKdV equation.

For the sake of brevity, we shall refer to mKdV equation as the potential mKdV equation henceforth
\begin{eqnarray}\label{pmkdv}
u_t=u_{xxx}+2u_x^3.
\end{eqnarray}
The generalized $K$ symmetries, denoted as $\tilde{K_{m}}$, are defined through the operator 
\begin{eqnarray}
\tilde{\Phi}\equiv \partial_x^2+4u_x\partial_x^{-1}u_x\partial_x.
\end{eqnarray}
The symmetries for different orders $m$ are then given by
\begin{eqnarray}\label{tkm}
\tilde{K_{m}}=\tilde{\Phi}^m u_x,~~m=1,2,...,\infty.
\end{eqnarray}
We explicitly compute the first few of these symmetries
\begin{eqnarray*}
\tilde{K_1}&=&u_x,\\
\tilde{K_2}&=&u_{xxx}+2u_x^3,\\
\tilde{K_3}&=&u_{5x}+10u_x(u_{xx}^2+u_xu_{xxx})+6u_x^5,\\
\tilde{K_4}&=&u_{7x}+14u_x^2u_{5x}+56u_xu_{xx}u_{4x}+14(5u_{xx}^2+3u_xu_{xxx}+5u_x^4)u_{xxx}+140u_x^3u_{xx}^2+20u_x^7,\\
\tilde{K_5}&=&u_{9x}+18u_x^2u_{7x}+108u_xu_{xx}u_{6x}+6(21u_x^4+38u_xu_{xxx}+35u_{xx}^2)u_{5x}+6(23u_xu_{4x}\\&+&168u_x^3u_{xx}+126u_{xx}u_{xxx})u_{4x}+182u_{xxx}^3+84(9u_xu_{xxx}+5u_x^4+37u_{xx}^2)u_x^2u_{xxx}\\&+&42(30u_x^4+19u_{xx}^2)u_xu_{xx}^2+70u_x^9.
\end{eqnarray*}
%By applying these symmetries, we are able to extract a series of solutions that describe multi-wave interactions within the framework of the potential mKdV equation. 

	To elaborate, consider the $n$-wave solution structure given by  
	\begin{equation}
	u = u_{nw} = u(\xi_1,\ \xi_2,\ \ldots, \ \xi_n),\ \xi_i \equiv k_i x + \omega_i t + c_i,\ i=1,\ 2,\ \ldots,\ n, \label{unw}
	\end{equation}
	where the wave profile $u$ is expressed as a function of phase variables $\xi_i$. Each $\xi_i$ is a linear combination of spatial and temporal components, modulated by arbitrary constants $k_i$ and $c_i$, along with frequency components $\omega_i$ that encapsulate the wave numbers and other parameters.

	Regarding the $n$-wave solution delineated by equation \eqref{unw}, it is imperative to recognize that the myriad $K$-symmetries \eqref{tkm} are meticulously constructed as linear combinations of the center translation symmetries, each denoted as $u_{c_i}=\partial_{c_i}u_{nw}$
	\begin{eqnarray}\label{KN}
	\tilde{K_m}|_{u=u_{nw}}=\sum_{i=1}^na_{mi}u_{c_i},~~ m=1,2,...,\infty.
	\end{eqnarray}
	Here, $a_{mi}$ are coefficients that determine the contribution of each symmetry to the wave entity, indexed by $i$, which spans from $1$ to $n$, attributing each symmetry to a respective wave component.

\subsection{Two-wave solution}

	In the wave dynamics, the two-wave solution is of paramount importance, particularly when we consider the number of waves, denoted by $n$, to be specifically 2. This scenario engenders a solution that is characterized by the function
\begin{eqnarray}\label{u2w}
u=u(\xi_1,\xi_2), ~\xi_i=k_ix+\omega_{i}t+c_i,
 ~i=1,2.
\end{eqnarray}
Within this context, it is crucial to recognize that the partial derivative of $u$ with respect to $c_i$ is equivalent to the derivative with respect to the phase variable $\xi_i$, expressed mathematically
\begin{eqnarray}\label{uci}
u_{c_i}=u_{\xi_i}.
\end{eqnarray}
By embedding equations \eqref{u2w} and \eqref{uci} into the $K$-symmetry equation \eqref{KN}, we derive a system of equations that governs the two-wave interaction. This system is given by
\begin{eqnarray}
&&\tilde{K_m}-\sum_{i=1}^2a_{mi}u_{\xi_i}=0,  ~m=1,...,4.
\end{eqnarray}
This set of equations is pivotal in elucidating the underlying two-wave solutions.
\subsubsection{Complexiton solution}
Upon solving the aforementioned equations, 
we get the two-wave complexiton solution of the potential mKdV equation \eqref{pmkdv} \cite{hbib14}
\iffalse
\begin{eqnarray}
u=\pm {\rm i} \ln\left(\frac{H_{2-}}{H_{2+}}\right).
\end{eqnarray}
The functions $H_{2+}$ and $H_{2-}$ are delineated as
\begin{eqnarray}
H_{2\pm}=\left({\rm e}^{2\xi_1}-{\rm e}^{2\xi_1+4\xi_2{\rm i}}\right)k_1\pm \left({\rm e}^{4\xi_1+2\xi_2{\rm i}}-{\rm e}^{2\xi_2\rm i}\right)k_2{\rm i}.
\end{eqnarray} 
\fi
\begin{equation}
u=\pm 2 {\rm i ~\arctanh}  \left(\frac{k_2\sinh(\xi_1)}{k_1\sin(\xi_2)}\right).
\end{equation}
The coefficients $a_{m1}$ and $a_{m2}$ are articulated through the following expression
\begin{eqnarray*}
	a_{m1}&=&\frac{1}{2}\left[( k_1+{\rm i }k_2)\left(\frac{(k_1^2+k_2^2)^2}{k_1^2-k_2^2-2k_1k_2{\rm i}}\right)^m+( k_1-{\rm i}k_2)\left(\frac{(k_1^2+k_2^2)^2}{k_1^2-k_2^2+2k_1k_2{\rm i}}\right)^m \right],\\
	a_{m2}&=&\frac{1}{2}\left[( k_2-{\rm i }k_1)\left(\frac{(k_1^2+k_2^2)^2}{k_1^2-k_2^2-2k_1k_2{\rm i}}\right)^m+( k_2+{\rm i}k_1)\left(\frac{(k_1^2+k_2^2)^2}{k_1^2-k_2^2+2k_1k_2{\rm i}}\right)^m \right].
\end{eqnarray*}

Furthermore, we identify a second kind of solution, which is expressed as  
\begin{equation}
u=\pm 2 {\rm i ~\arctanh}  \left(\frac{k_2\sin(\xi_1)}{k_1\cos(\xi_2)}\right),
\end{equation}
with the corresponding coefficients given by
\begin{eqnarray*}
a_{m1}&=&\frac{(-1)^m}{2}\left[(k_1-k_2)^{1+2m}+(k_1+k_2)^{1+2m}\right],\\
a_{m2}&=&\frac{(-1)^{m}}{2}\left[(k_1+k_2)^{1+2m}-(k_1-k_2)^{1+2m}\right].
\end{eqnarray*}
This solution represents a distinct wave pattern, highlighting the oscillatory nature of the interaction between the two waves, with the sine and cosine functions capturing the periodic behavior of the waves.

The third kind of solution is articulated as 
	\begin{equation}
	u=\pm 2 {\rm i ~\arctanh}  \left(\frac{k_2 \xi_1}{k_1\cos(\xi_2)}\right).
	\end{equation}
The coefficients for this solution are delineated as
	\begin{eqnarray*}
		a_{m1}&=&(-1)^mk_1^{1+2m},\\
		a_{m2}&=&(-1)^{m}(1+2k)k_1^{2m}k_2.
	\end{eqnarray*}
This solution indicates a more straightforward interaction where the amplitude of one wave is directly influenced by the other.

\subsubsection{Breather solution}
Beyond the complexiton solution, the two-wave scenario also accommodates breather solutions. one kind of breather solution \cite{mbib24,dbib8}, for instance
\begin{eqnarray}
u=\pm 2\arctan\left(\frac{k_2\sin(\xi_1)}{k_1\cosh(\xi_2)}\right)
\end{eqnarray}
\iffalse
\begin{eqnarray}
u=\pm {\rm i}\ln\left(\frac{K_{2-}}{K_{2+}}\right)
\end{eqnarray}
\begin{eqnarray}
K_{2\pm}=\left({\rm e}^{\xi_1i}+{\rm e}^{2\xi_2+\xi_1i}\right)k_1\pm \left({\rm e}^{\xi_2}-{\rm e}^{2\xi_1i+\xi_2}\right)k_2
\end{eqnarray}
\fi
describes a localized wave packet that maintains its shape as it propagates, a phenomenon known as a breather. This is a direct manifestation of the nonlinear nature of the system, where the waves interact in such a way as to create a stable, localized disturbance.
The coefficients $a_{m1}$ and $a_{m2}$ for the breather solution are articulated by
\begin{eqnarray*}
	a_{m1}&=&\frac{(-1)^{k+1}}{2}\left[-( k_1+{\rm i}k_2)\left(\frac{(k_1^2+k_2^2)^2}{k_1^2-k_2^2-2k_1k_2{\rm i}}\right)^m+({\rm i} k_2-k_1)\left(\frac{(k_1^2+k_2^2)^2}{k_1^2-k_2^2+2k_1k_2{\rm i}}\right)^m \right],\\
	a_{m2}&=&\frac{(-1)^{k+1}}{2}\left[( {\rm i}k_1-k_2)\left(\frac{(k_1^2+k_2^2)^2}{k_1^2-k_2^2-2k_1k_2{\rm i}}\right)^m-( k_2+{\rm i}k_1)\left(\frac{(k_1^2+k_2^2)^2}{k_1^2-k_2^2+2k_1k_2{\rm i}}\right)^m \right].
\end{eqnarray*}

	\subsubsection{Soliton solution}
	The two soliton solution, on the other hand
\begin{eqnarray}
u=\pm 2 \arctan\left(\frac{k_2\sinh(\xi_1)}{k_1\cosh(\xi_2)}\right)
\end{eqnarray}
represents a stable wave soliton that maintains its shape and speed even after interacting with other waves. 
The coefficients are articulated as 
\begin{eqnarray}
a_{m1}&=&\frac{(k_1-k_2)^{1+2m}}{2}+\frac{(k_1+k_2)^{1+2m}}{2},\\
a_{m2}&=&\frac{(k_1+k_2)^{1+2m}}{2}-\frac{(k_1-k_2)^{1+2m}}{2}.
\end{eqnarray}

\subsubsection{Double pole solution}
The double pole solution \cite{mbib25,zbib23}, representing a soliton-antisoliton pair traveling in the same direction, is given by
\begin{eqnarray}
u=\pm 2\arctan\left(\frac{k_2\xi_1}{k_1\cosh(\xi_2)}\right).
\end{eqnarray}
The coefficients for this solution are 
\begin{eqnarray*}
a_{m1}&=&(1+2k)k_1k_2^{2m},\\
a_{m2}&=&k_2^{1+2m}.
\end{eqnarray*}

\subsection{Three-wave solution}
When the number of waves, $n$, is equal to three, the three-wave solution is given by
\begin{eqnarray}\label{u3s}
u=u(\xi_1,\xi_2,\xi_3),~ \xi_i=k_ix+\omega_{i}t+c_i, ~u_{c_i}=u_{\xi_i},~ i=1,2,3.
\end{eqnarray}
By substituting equations \eqref{u3s} into \eqref{KN}, we arrive at a system of equations for the $K$-symmetries
\begin{eqnarray}
&&\tilde{K_m}-\sum_{i=1}^3a_{mi}u_{\xi_i}=0,  ~m=1,...,6.
\end{eqnarray}
Solving above system, we get three-soliton solution of the potential mKdV equation
\begin{eqnarray}
u(\xi_1,\xi_2,\xi_3)=\pm{\rm i}\ln\left(\frac{G_{3-}}{G_{3+}}\right).
\end{eqnarray}
The coefficients $a_{mi}$ and the functions $G_{3+}$ and $G_{3-}$ are defined as follows
\begin{eqnarray*}
a_{mi}&=&k_{i}^{2m+1},\\
G_{3\pm}&=&1\pm{\rm e}^{\xi_1}\pm{\rm e}^{\xi_2}\pm{\rm e}^{\xi_3}+\frac{(k_1-k_2)^2}{(k_1+k_2)^2}{\rm e}^{\xi_1+\xi_2}+\frac{(k_1-k_3)^2}{(k_1+k_3)^2}{\rm e}^{\xi_1+\xi_3}+\frac{(k_2-k_3)^2}{(k_2+k_3)^2}{\rm e}^{\xi_2+\xi_3}\\&\pm&\frac{(k_1-k_2)^2(k_1-k_3)^2(k_2-k_3)^2}{(k_1+k_2)^2(k_1+k_3)^2(k_2+k_3)^2}{\rm e}^{\xi_1+\xi_2+\xi_3}.
\end{eqnarray*}

In the pursuit of the $n$-soliton solution, we commence with the system of equations emanating from the $K$-symmetries
\begin{eqnarray}
&&\tilde{K_m}-\sum_{i=1}^na_{mi}u_{\xi_i}=0,~~m=1,...,2n,
\end{eqnarray}
where the coefficients are defined as $a_{m,i}=k_i^{2m+1}$.
Upon resolving this system, we derive the $n$-soliton solution of the potential mKdV equation \eqref{pmkdv}
\begin{eqnarray}
u_{ns}=\pm {\rm i} \ln\left(\frac{G_{n-}}{G_{n+}}\right).
\end{eqnarray}
The function $G_{n+}$ and $G_{n-}$ are articulated as
\begin{eqnarray}
G_{n\pm}=\sum_{\mu=1,0}(\pm 1)^{\sum_{i=1}^n\mu_i}\exp\left[\sum_{i=1}^{n}\mu_i\xi_i+\sum_{1\leq i<l}^n\mu_i\mu_l\theta_{il}\right]
\end{eqnarray}
with the exponential term  $\exp(\theta_{jl})$ specified by
\begin{eqnarray}
\exp(\theta_{jl})=\left(\frac{k_j-k_l}{k_j+k_l}\right)^2.
\end{eqnarray}
The summation over $\mu$ encompasses all possible permutations of $\mu_i = 0, 1$ for $i = 1, 2, \ldots, n$.

%\section{nonlocal symmetry of linear superposition}

\section{Conclusions and discussions}
The present study concludes with some advancements in the understanding of the mKdV equation, particularly in the context of its symmetries and multi-wave solutions. By examining the infinite many symmetries inherent to the mKdV equation, we have shed light on the physical interpretations of these symmetries.
Our analysis has demonstrated that the infinitely many symmetries of the mKdV equation are intricately linked to the translational invariance of its multi-wave solutions. Specifically, within the confines of a fixed $n$-soliton solution, we have observed that the ostensibly infinite many $K$-symmetries and $\tau$-symmetries are effectively degenerated to a finite set comprising $2n$ independent symmetries, namely, the $K$-symmetries and $\tau$-symmetries can be expressed as linear combinations of the translation symmetries associated with the wave parameters. 
This finding challenges the notion of completeness in the realm of infinitely many known symmetries, suggests that the mKdV equation, and potentially other integrable systems, possess a more extensive symmetry structure than previously recognized and necessitates a refined perspective on identifying and classifying these symmetries. This insight may lead to the discovery of new solutions and a deeper understanding of the physical phenomena described by these systems.

Furthermore, our finding also implies that the application of symmetry-based method could lead to the identification of novel exact solutions within the mKdV equation and potentially within other integrable systems. Our investigation has unveiled the existence of various categories of $n$-wave solutions, encompassing $n$-soliton solutions, multiple breathers, complexitons, and $n$-periodic wave solutions. By examining the potential mKdV equation, we have derived the two-wave complexiton solution, breather solution, double pole and the three-soliton solution.

In conclusion, this work has provided a deeper understanding of multi-wave solutions and their symmetries within integrable systems. Future research directions may include exploring the potential applications of these findings, as well as continuing to investigate the completeness and classification of symmetries in integrable systems.

\section*{Acknowledgement}
The work was sponsored by the National Natural Science Foundations of China (Nos. 12301315,  12235007, 11975131), the Natural Science Foundation of Zhejiang Province No. LQ20A010009.

%\normalem
%\bibliographystyle{unsrt}
%\bibliographystyle{plain}
\bibliographystyle{elsarticle-num}
\bibliography{ref}

\end{CJK*}
\end{document}